%1234567890123456789012345678901234567890123456789012345678901234567890
\input phyzzx
%square (in math mode)
\def\sqr#1#2{{\vcenter{\hrule height.#2pt
      \hbox{\vrule width.#2pt height#1pt \kern#1pt
          \vrule width.#2pt}
      \hrule height.#2pt}}}

\def\square{{\mathchoice{\sqr84}{\sqr84}{\sqr{5.0}3}{\sqr{3.5}3}}}
%
%%%%%%%%%%%%%%%%%%%%%%%%%%%%%%%%%%%%%%%%%%%%%%%%%%%%%%%%%%%%%%%%%%%%%%
\REF\KAW{
H. Kawai and M. Ninomiya,
{\sl Nucl. Phys.\/} {\bf B336} (1990) 115;\hfil\break
H. Kawai, Y. Kitazawa and M. Ninomiya,
{\sl Nucl. Phys.\/} {\bf B393} (1993) 280.}
\REF\TAN{
Y. Tanii, S. Kojima and N. Sakai,
{\sl Phys. Lett.\/} {\bf B322} (1994) 59;\hfil\break
S. Kojima, N. Sakai and Y. Tanii,
{\sl Int. J. Mod. Phys.\/} {\bf A9} (1994) 5415.}
\REF\WEI{
S. Weinberg, {\sl in\/} General Relativity, an Einstein Centenary
Survey, ed.\ S. W. Hawking and W. Israel (Cambridge University Press,
1979)}
\REF\GAS{
R. Gastmans, R. Kallosh and C. Truffin,
{\sl Nucl. Phys.\/} {\bf B133} (1978) 417;\hfil\break
S. M. Christensen and M. J. Duff,
{\sl Phys. Lett.\/} {\bf B79} (1978) 213.}
\REF\JAC{
I. Jack and D. R. T. Jones,
{\sl Nucl. Phys.\/} {\bf B358} (1991) 695.}
\REF\KIT{
H. Kawai, Y. Kitazawa and M. Ninomiya,
{\sl Nucl. Phys.\/} {\bf B404} (1993) 684;\hfil\break
T. Aida, Y. Kitazawa, H. Kawai and M. Ninomiya,
{\sl Nucl. Phys.\/} {\bf B427} 1994 158;\hfil\break
Y. Kitazawa,
{\sl Nucl. Phys.\/} {\bf B453} (1995) 477.}
\REF\SAK{
S. Kojima, N. Sakai and Y. Tanii,
{\sl Nucl. Phys.\/} {\bf B426} (1994) 223;\hfil\break
{\sl Mod. Phys. Lett.\/} {\bf A10} (1995) 2391.}
\REF\ODI{
E. Elizalde and S. D. Odintsov,
{\sl Phys. Lett.\/} {\bf B313} (1993) 347;\hfil\break
{\sl Phys. Lett.\/} {\bf B347} (1995) 211;\hfil\break
{\sl Mod. Phys. Lett.\/} {\bf A10} (1995) 2001.}
\REF\REC{
Y. Kikukawa and K. Yamawaki,
{\sl Phys. Lett.\/} {\bf B234} (1990) 497;\hfil\break
J. Jinn-Zustin,
{\sl Nucl. Phys.\/} {\bf B367} (1991) 105;\hfil\break
S. Hands, A. Koci\'c and J. B. Kogut,
{\sl Phys. Lett.\/} {\bf B273} (1991) 111;\hfil\break
H.-J. He, Y.-P Kuang, Q. Wang and Y.-P. Yi,
{\sl Phys. Rev.\/} {\bf D45} (1992) 4610; and references therein.}
\REF\WIL{
K. G. Wilson,
{\sl Phys. Rev.\/} {\bf D7} (1973) 2911.}
\REF\GRO{
D. J. Gross and A. Neveu,
{\sl Phys. Rev.\/} {\bf D10} (1979) 3235.}
\REF\NAM{
Y. Nambu and G. Jona-Lasinio,
{\sl Phys. Rev.\/} {\bf 122} (1961) 345.}
\REF\GROSS{
D. J. Gross, {\sl in\/} Methods in field theory, Les Houches 1975,
ed.\ R. Balian and J. Zinn-Justin (North-Holland, Amsterdam, 1976)} 
\REF\ROS{
B. Rosenstein and B. J. Warr,
{\sl Phys. Rev. Lett\/} {\bf 62} (1989) 1433.}
%
%%%%%%%%%%%%%%%%%%%%%%%%%%%%%%%%%%%%%%%%%%%%%%%%%%%%%%%%%%%%%%%%%%%%
%
\overfullrule=0pt
\pubnum={IU-MSTP/6; hep-th/9601061}
\date={January 1996}
\titlepage
\title{Renormalization Group in $2+\epsilon$ Dimensions and
$\epsilon\to2$:\break A simple model analysis}
\author{
Nobuaki Nagao and Hiroshi Suzuki\foot{
e-mail: hsuzuki@mito.ipc.ibaraki.ac.jp}}
\address{
Department of Physics, Ibaraki University, Mito 310, Japan}
\abstract{
Using a simple solvable model, i.e., Higgs--Yukawa system with an
infinite number of flavors, we explicitly demonstrate how a
dimensional continuation of the $\beta$ function in two
dimensional MS scheme {\it fails\/} to reproduce the correct behavior
of the $\beta$ function in four dimensions. The mapping between
coupling constants in two dimensional MS scheme and a conventional
scheme in the cutoff regularization, in which the dimensional
continuation of the $\beta$ function is smooth, becomes singular when
the dimension of spacetime approaches to four. The existence of a
non-trivial fixed point in $2+\epsilon$ dimensions continued to four
dimensions $\epsilon\to2$ in the two dimensional MS scheme is
spurious and the asymptotic safety cannot be imposed to this model
in four dimensions.}
\endpage
%%%%%%%%%%%%%%%%%%%%%%%%%%%%%%%%%%%%%%%%%%%%%%%%%%%%%%%%%%%%%%%%%%%%%%%%%%

Much interest on quantum gravity in $2+\epsilon$ dimensions has
recently been revived [\KAW,\TAN]. This dimensional continuation
approach to {\it four\/} dimensional quantum gravity was originally
proposed by Weinberg [\WEI], in a connection with a possible way to
give a predictive power on non-renormalizable theories, called
``asymptotic safety.''

Einstein gravity in four dimensions is power counting
non-renormalizable and thus one needs an infinite number of counter
terms (and coupling constants) to remove the ultraviolet (UV)
divergences. By the asymptotic safety, one requires all the
renormalized coupling constants flow into (or remain at) a certain
fixed point in the UV limit by the renormalization group (RG).
Imposing this condition, the Landau singularity or the UV renormalon
is avoided. The condition moreover puts the renormalized parameters
on a finite dimensional surface, called UV critical surface [\WEI],
and gives a predictive power to the theory; all the infinite
coupling constants are parameterized by a finite number of coupling
constants on the surface.

The possibility whether the asymptotic safety can be imposed or not
hence depends on the existence of the fixed point of RG. Since the
mass dimension of the gravitational constant is negative for $D>2$,
to have a non-trivial theory (namely $G_R\neq0$), we have to find
a non-trivial fixed point of RG.

It is of course in practice impossible to determine the full
structure of the RG flow in the infinite dimensional coupling
constant space. The idea of Weinberg is, however, that if one
uses MS scheme in the dimensional regularization, the task to find a
non-trivial fixed point is drastically simplified. Namely if one
could find some non-trivial zero of the $\beta$ function in this
scheme by setting all the non-renormalizable (as well as super
renormalizable) interactions zero, then it is a {\it true\/} fixed
point in the infinite dimensional coupling constant space. This is a
peculiarity of the dimensional regularization [\WEI].

From this viewpoint, it is natural to take the two dimensional
MS scheme (MS2), because Einstein gravity is power counting
renormalizable in $D=2$ and we are interested in a non-trivial theory
$G_R\neq0$. The result can then be continued to $2+\epsilon$
dimensions as long as $\epsilon$ is an irrational number, because
the dimensional regularization puts the Feynman integral in a
form that is singular only at rational dimensions; MS2 gives
a {\it renormalization\/} in $2+\epsilon$ dimensions.

Now according to the actual one loop calculation of the $\beta$
function of the gravitational constant [\GAS] in this scheme,
there exists a non-trivial zero. Therefore
at least for $\epsilon$ small
enough (in the spirit of the $\epsilon$ expansion) the
asymptotic safety may be imposed to Einstein gravity.

Although it has now been realized [\JAC,\KIT,\SAK,\ODI]
that the original
program of Weinberg as it stands does not work due to a peculiarity of
the Einstein action (i.e., the kinematical pole), this fact may even
force us to modify the Einstein action [\KIT,\SAK,\ODI].

Besides this problem of the Einstein action, really to reach four
dimensions in this approach, it is a crucial if certain
properties obtained in $2+\epsilon$ dimensions can smoothly be
continued to $\epsilon\to2$. Among of them, the existence of a
non-trivial fixed point and the eigenvalue of RG flow at the fixed
point (the critical exponents) should smoothly be continued.
This is crucial to conclude the possibility to impose the asymptotic
safety in four dimensions (in which we {\it are\/} interested).

The aim of this article is to address the above point: Suppose
MS2 shows a fixed point in $2+\epsilon$ dimensions. They can
be continued to $\epsilon\to2$?

To study this point without any {\it ad hoc\/} approximation, we will
consider a simple exactly calculable model, i.e., Higgs--Yukawa system
with an infinite number of flavors:\foot{Of course this model, or its
analogue, i.e., the four fermi interaction in $D$ dimensions, have
been studied countless times in the literature. For relatively recent
articles, see [\REC].}
$$
   {\cal L}=
   {N\over2}Z\partial_\mu\sigma\partial^\mu\sigma-{N\over4G}\sigma^2
   -{N\over4!}g\sigma^4+\bar\psi i\gamma^\mu\partial_\mu\psi
   -\bar\psi\sigma\psi,
\eqn\one
$$
where $\psi$ is an $N$ component column vector,
$\psi=(\psi_1,\cdots,\psi_N)^T$ and $Z$, $G$ and $g$ are bare
coupling constants. We shall consider $N\to\infty$ limit of this
system (the leading order of $1/N$ expansion [\WIL,\GRO]) in $D$
dimensional spacetime.

In what follows we shall find that the answer to the above question is
negative: The fixed point in $2+\epsilon$ dimensions continued
to $\epsilon\to2$, obtained in MS2 is spurious. The asymptotic
safety cannot be imposed on this model.\foot{One should be curious
on this statement because the lagrangian \one\ is renormalizable for
$D\leq4$ and it is not really necessary to rely on the asymptotic
safety. Our point is, however, to demonstrate how the dimensional
continuation of MS2 to four dimensions fails using this renormalizable
model.}

Now in \one\ radiative corrections due to the scalar field $\sigma$
is trivial because $N$ is just the inverse of the Plank constant and
the classical contribution dominates for $N\to\infty$ (tree level
exact). On the other hand for the fermion field $\psi$, the one loop
correction is exact. So we can compute any Green functions in a
closed form (even implicit) in this large $N$ limit [\WIL,\GRO]. Note
that in \one\ we have normalized the Yukawa coupling unity and
instead introduced the wave function normalization factor $Z$.
Throughout this article, we will use a convention $\tr1=2^{D/2}$.

The central physics of the model \one\ is of course the dynamical
chiral symmetry ($\psi\to\gamma_5\psi$, $\sigma\to-\sigma$ in even
dimensions) breaking [\GRO,\NAM]. To detect this, one may shift the
scalar field $\sigma\to\sigma+m$ and impose the vanishing of the
tadpole diagram [\NAM], which is calculated in the dimensional
regularization as:
$$
   \Gamma^{(1)}_\sigma/N=
   -{m\over2G}-{gm^3\over3!}+{\Gamma(1-D/2)\over(2\pi)^{D/2}}m^{D-1}=0.
\eqn\two
$$
One may discuss the dynamical symmetry breaking by taking $m\neq0$
solution of \two. In this article, however, we will only consider the
symmetric solution $m=0$ for a simplicity of presentation, because
we are interested in the UV behavior of the system. Actually, by
taking an appropriate renormalization condition it can be checked
that the same counter terms in the symmetric phase can remove the UV
divergences in the breaking phase.

In the dimensional regularization, the two point 1PI function of
$\sigma$ is given by
$$
   \Gamma^{(2)}_\sigma(p)/N=Zp^2-{1\over2G}
   +{\Gamma(D/2)^2\Gamma(1-D/2)\over2(2\pi)^{D/2}\Gamma(D-1)}
    (-p^2)^{D/2-1},
\eqn\three
$$
and the four point function is given by
$$
\eqalign{
   &\Gamma^{(4)}_\sigma(p_1,p_2,p_3)/N
\cr
   &=-g-{D(D+2)\Gamma(2-D/2)\over4(2\pi)^{D/2}}
   \int_0^1dz\,z^2\int_0^1dy\,y\int_0^1dx\,
   [f(p_1,p_2,p_3;x,y,z)]^{D/2-2}
\cr
   &\qquad+({\rm five\ permutations\ on\ }p_1, p_2, p_3)
    +({\rm no\ pole\ part}),
\cr
}
\eqn\four
$$
where
$$
\eqalign{
   f(p_1,p_2,p_3;x,y,z)&=-(1-xyz)p_1^2-(1-yz)p_2^2-(1-z)p_3^2
\cr
   &\qquad-2(1-yz)p_1\cdot p_2-2(1-z)p_1\cdot p_3-2(1-z)p_2\cdot p_3.
\cr
}
\eqn\five
$$
Higher point 1PI functions $\Gamma^{(n)}_\sigma(p_1,\cdots,p_{n-1})$
with $n\geq6$ have no pole for $D\leq4$ and so are irrelevant for the
following discussion.

In the large $N$ limit, any Green function among $\psi$ is given by
tree diagrams made from vertices
$\Gamma^{(n)}_\sigma(p_1,\cdots,p_{n-1})$
and the propagator of $\sigma$ field (and putting $\bar\psi\psi$ to
the external lines). All the information of the UV divergence is thus
contained in the above two functions.

\section{MS scheme in $D=2$}
Let us start our discussion on RG from the two dimensional MS scheme
(MS2). Setting $D=2+\epsilon$, only pole terms $1/\epsilon^n$ are
subtracted in this scheme [\WEI]. We define the renormalization
constants (note the canonical dimension of $\sigma$ is $1$
irrespective of the spacetime dimension $D$) as
$$
   Z=\mu^{-2+\epsilon}Z_ZZ_R,\quad
   G=\mu^{-\epsilon}Z_GG_R,\quad
   g=\mu^{-2+\epsilon}Z_gg_R,
\eqn\six
$$
from \three\ and \four, we see $Z_Z=Z_g=1$ and
$$
   Z_G={1\over1-{2\over\pi}{1\over\epsilon}G_R}
   =\sum_{n=0}^\infty\left({2\over\pi}\right)^n{1\over\epsilon^n}G_R^n.
\eqn\seven
$$
We stress that a pole singularity at $D=4$ in \three\ and
\four\ is not subtracted in this scheme.

From \seven, the $\beta$ functions are given by
$$
\eqalign{
   &\beta_Z\equiv\mu{\partial Z_R\over\partial\mu}=(2-\epsilon)Z_R,
   \quad\beta_G\equiv\mu{\partial G_R\over\partial\mu}
   =\epsilon G_R-{2\over\pi}G_R^2,
\cr
   &\beta_g\equiv\mu{\partial g_R\over\partial\mu}=(2-\epsilon)g_R.
\cr
}
\eqn\eight
$$
The $\beta$ function of $Z_R$ and $g_R$ is of course trivial in this
scheme. On the other hand the $\beta$ function of $G_R$ has
non-trivial zero [\WIL,\GRO] at $G_R^*=\pi\epsilon/2$ and so for
$\epsilon\neq0$ we have two fixed points in the full coupling constant
space:
$$
\eqalign{
   &Z_R=0,\quad G_R=0,\quad g_R=0,\qquad{\rm (A)},
\cr
   &Z_R=0,\quad G_R=G_R^*={\pi\epsilon\over2},\quad g_R=0,
   \qquad{\rm (B)}.
\cr
}
\eqn\nine
$$
For $\epsilon\leq2$ the point (A) is infrared (IR) stable. For
$\epsilon=2$ the direction of $Z_R$ and $g_R$ becomes scale invariant
and an arbitrary value of $Z_R$ and $g_R$ (with $G_R=0$ or $G_R^*$)
gives the fixed point.

In general a fixed point is characterized by the eigenvalue of a
matrix $\partial\beta_i/\partial g_j$ at the fixed point (critical
exponents) [\GROSS]:
$$
\eqalign{
   &2-\epsilon=4-D,\quad\epsilon=D-2,\quad2-\epsilon=4-D,\quad
   {\rm for\ (A)},
\cr
   &2-\epsilon=4-D,\quad-\epsilon=2-D,\quad2-\epsilon=4-D,\quad
   {\rm for\ (B)},
\cr
}
\eqn\ten
$$
and should be the {\it same\/} under the change of the
renormalization scheme [\GROSS]. Therefore they can be used to
identify the corresponding fixed points between different schemes.

Now according to \nine\ and \ten\ it seems that we may impose the
asymptotic safety by setting $Z_R=0$ and $g_R=0$. Note that \nine\ is
not $\epsilon$ expansion but is exact [\WIL]. After imposing the
asymptotic safety, the theory becomes the Gross--Neveu model [\GRO]
in $2+\epsilon$ dimensions. But there is no obstruction in \nine\
and \ten\ to take a limit $\epsilon\to2$ and
to go to the four dimensions.
Then $G_R\to\pi$ in the UV limit and we have a non-trivial theory in
four dimensions. This may be taken as a possible definition of a
four dimensional Gross--Neveu model.

As will be shown shortly (or as is easily expected) this is
not the case. The fixed point (B) for $\epsilon\to2$ is an artifact
of the present scheme. We will show this fact by finding a mapping
between the renormalized couplings in MS2 and a conventional scheme in
the cutoff regularization. However before going into this, let us
summarize what is actually happening in four dimensions.

\section{MS scheme in $D=4$}
To see the situation in four dimensions, it is most convenient to use
the four dimensional MS scheme (MS4). Setting $D=4-2\epsilon'$ and
$$
   Z=\mu^{-2\epsilon'}Z_ZZ_R,\quad G=\mu^{-2+2\epsilon'}Z_GG_R,\quad
   g=\mu^{-2\epsilon'}Z_gg_R,
\eqn\eleven
$$
we see the following choice removes the pole $1/\epsilon'$ in
\three\ and \four:
$$
   Z_Z=1-{1\over8\pi^2}{1\over\epsilon'}{1\over Z_R},\quad
   Z_G=1,\quad
   Z_g=1-{3\over2\pi^2}{1\over\epsilon'}{1\over g_R}.
\eqn\twelve
$$
Note that $G$ receives no radiative correction. From this we have the
$\beta$ functions in this scheme
$$
   \beta_Z=2\epsilon'Z_R-{1\over4\pi^2},\quad
   \beta_G=(2-2\epsilon')G_R,\quad
   \beta_g=2\epsilon'g_R-{3\over\pi^2}.
\eqn\thirteen
$$
The $\beta$ functions for $Z_R$ and $g_R$ seem peculiar but are
consistent. Remind
that we have taken the Yukawa coupling in \one\ unity. If we put
the Yukawa coupling $\lambda$ instead, we would have $\lambda^2$
and $\lambda^4$
in the second terms; or if one prefers
a standard form of the $\beta$
function, the inverse coupling may be considered
$$
   \mu{\partial Z_R^{-1}\over\partial\mu}
   =-2\epsilon'Z_R^{-1}+{1\over4\pi^2}Z_R^{-2}.
\eqn\fourteen
$$
Clearly for $\epsilon'=0$, $Z_R\to\infty$ as $\mu\to\infty$ (note
$Z_R^{-1}\to+\infty$ is connected to $Z_R^{-1}\to-\infty$).

From the form of the $\beta$ function \thirteen\ we see there exists
a unique fixed point,
$$
   Z_R={1\over8\pi^2}{1\over\epsilon'},\quad
   G_R=0,\quad
   g_R={3\over2\pi^2}{1\over\epsilon'},
\eqn\fifteen
$$
which is IR stable for $\epsilon'>0$. The corresponding
critical exponents read,
$$
  2\epsilon'=4-D,\quad
  \quad2-2\epsilon'=D-2,\quad
  \quad2\epsilon'=4-D.
\eqn\sixteen
$$
Comparing the critical exponents \sixteen\ and \ten, we realize that
the fixed point in \fifteen\ corresponds to (A) in \nine\ in MS2.
Where is the another fixed point (B)? By comparing the $\beta$
functions in two schemes \eight\ and \thirteen, (identifying both
of the renormalization scales $\mu$), we find a mapping between
$G_R^{\rm MS2}$ and $G_R^{\rm MS4}$ in $2<D<4$, (for which
{\it both\/} of schemes give a {\it renormalization}):
$$
   G_R^{\rm MS4}={G_R^{\rm MS2}\over
                  1-{2\over\pi\epsilon}G_R^{\rm MS2}}.
\eqn\seventeenone
$$
We realize the fixed point (B) in \nine, $G_R^{\rm MS2}=\pi\epsilon/2$
is mapped to the infinity of $G_R^{\rm MS4}$ for {\it arbitrary\/}
dimension $2<D<4$. The mapping from MS2 to MS4 is therefore
singular at the fixed point (B) in MS2.

At $D=4$ there is {\it no\/} fixed point from \thirteen\ and it is
definitely impossible to impose the asymptotic safety. Moreover in UV
limit $Z_R$ and $g_R$ diverge for any choice of the bare parameter,
so the theory is pathological in four dimensions.\foot{This depends on
what is called the ``coupling constant.'' By rescaling
$\sigma\to\sigma/\sqrt{Z}$, the Yukawa coupling becomes $1/\sqrt{Z}$.
Then RG tells that the theory becomes weakly interacting massless boson
and fermion, which is not pathological at all. Our definition of the
coupling constant is motivated by the Gross--Neveu model, in which
the fermion has a hard contact interaction. See also a discussion in
Conclusion.}

In conclusion, the non-trivial fixed point in four dimensions, that
is detected by a dimensional continuation of MS2 is spurious. To
see this point much clearer, we shall consider one more
another scheme in the
next section, in which the $\beta$ function is continuous in the whole
range of the dimension $2\leq D\leq4$. In some sense
it interpolates the two
dimensional MS scheme and the four dimensional MS scheme.

\section{Cutoff regularization}
We apply the Euclidean momentum cutoff regularization in this section
($k^4\equiv k^0/i$), putting the momentum cutoff $\Lambda$
after a symmetrization of a denominator of the Feynman integral. The
two point 1PI function in $D$ dimensions in this regularization reads
$$
   \Gamma^{(2)}_\sigma(p)/N=Zp^2-{1\over2G}
   -{1\over(2\pi)^{D/2}\Gamma(D/2)}
   \int_0^1dx\int_0^{\Lambda^2}ds\,s^{D/2-1}
   {-p^2x(1-x)-s\over[-p^2x(1-x)+s]^2}.
\eqn\seventeen
$$
The four point function at the symmetric point reads
$$
\eqalign{
   &\Gamma^{(4)}_\sigma(p_1,p_2,p_3)/N
    \bigr|_
   {p_i\cdot p_j=-\mu^2\delta_{ij}+{1\over3}\mu^2(1-\delta_{ij})}
\cr
   &=-g-{36\over(2\pi)^{D/2}\Gamma(D/2)}
   \int_0^1dx\int_0^1dy\,y\int_0^1dz\,z^2
   \int_0^{\Lambda^2}ds\,s^{D/2+1}[s+\mu^2g(x,y,z)]^{-4}
\cr
   &\quad+({\rm finite\ part}),
\cr
}
\eqn\eighteen
$$
where
$$
   g(x,y,z)=z(1-z+{1\over3}y-{1\over3}xy+{2\over3}yz
              +{2\over3}xyz+{2\over3}xy^2z-y^2z-x^2y^2z).
\eqn\nineteen
$$
Under this regularization, we take the following Gell-Mann--Low type
renormalization condition:
$$
\eqalign{
  &\Gamma^{(2)}_\sigma(p)/N\bigr|_{p^2=-\mu^2}
       =-\mu^{D-2}Z_R-{\mu^{D-2}\over2G_R},
\cr
  &(\partial/\partial p^2)\Gamma^{(2)}_\sigma(p)/N\bigr|_{p^2=-\mu^2}
       =\mu^{D-4}Z_R,
\cr
   &\Gamma^{(4)}_\sigma(p_1,p_2,p_3)/N
    \bigr|_{p_i\cdot p_j=
                 -\mu^2\delta_{ij}+{1\over3}\mu^2(1-\delta_{ij})}
    =-\mu^{D-4}g_R.
\cr
}
\eqn\twenty
$$

Now the functions in \eighteen\ and \seventeen\ have different
divergent behavior depending on the spacetime dimension. Therefore
$2<D<4$ case and $D=4$ case should separately be treated.

For $2<D<4$, we first take a derivative $\partial/\partial p^2$
of the both sides of \seventeen\ and compare it with the second of
\twenty. We see (for $\Lambda\to\infty$)
$$
   Z=\mu^{D-4}\left[Z_R
         -{(D-1)\Gamma(D/2)^2\Gamma(2-D/2)\over(2\pi)^{D/2}\Gamma(D)}
         \right].
\eqn\twentyone
$$
Of course there is no divergence here and this gives the $\beta$
function of $Z_R$ in $2<D<4$,
$$
   \beta_Z=(4-D)Z_R
   -{2(D-1)\Gamma(D/2)^2\Gamma(3-D/2)\over(2\pi)^{D/2}\Gamma(D)}.
\eqn\twentytwo
$$

Similarly a comparison of \eighteen\ and \twenty\ gives
$$
\eqalign{
   g=\mu^{D-4}\Biggl[&g_R
     -{6\Gamma(2+D/2)\Gamma(2-D/2)\over(2\pi)^{D/2}\Gamma(D/2)}
      \int_0^1dx\int_0^1dy\,y\int_0^1dz\,z^2h(x,y,z)^{D/2-2}
\cr
   &\quad+O((4-D)^0)\Biggr],
\cr
}
\eqn\twentythree
$$
since the finite part has no singularity at $D=4$. The $\beta$
function of $g_R$ is therefore given by
$$
   \beta_g=(4-D)g_R-{3\over\pi^2}+O((4-D)).
\eqn\twentyfour
$$

To obtain the $\beta$ function of $G_R$, we directly take a derivative
$\mu\partial/\partial\mu$ of the both sides of the first of \twenty\
and \seventeen. After noting \twentyone\ and \twentytwo, we have for
$2<D<4$,\foot{This is {\it not\/} the same as RG function in
Gross--Neveu model in $2<D<4$ dimensions [\WIL,\GRO,\REC] because the
introduction of $Z$ modifies the divergent structure of $G$.
If we start $Z=g=0$ in \one\ instead (the Gross--Neveu model) we will
have $1/(D-4)$ pole in the second term.}
$$
   \beta_G=(D-2)G_R
   -{4(D-1)\Gamma(D/2)^2\Gamma(3-D/2)\over(2\pi)^{D/2}\Gamma(D)}G_R^2.
\eqn\twentyfive
$$

Repeating all the above steps in $D=4$, we have
$$
   Z=Z_R-{1\over8\pi^2}\ln\Lambda^2/\mu^2,\quad
   g=g_R-{3\over2\pi^2}\ln\Lambda^2/\mu^2+{\rm const.},
\eqn\twentysix
$$
and thus
$$
   \beta_Z=-{1\over4\pi^2},\quad\beta_g=-{3\over\pi^2}.
\eqn\twentyseven
$$
On the other hand,
$$
   \beta_G=2G_R-{1\over2\pi^2}G_R^2.
\eqn\twentyeight
$$

We note that although the divergence structure is different for
$2<D<4$ and $D=4$ (for example $Z$ is finite in $2<D<4$ but is
logarithmically divergent for $D=4$), the $\beta$ functions themselves
are {\it continuous\/} in this scheme. Namely the $\beta$ functions in
\twentytwo, \twentyfour\ and \twentyfive\ have a correct $D\to4$
limit, \twentyseven\ and \twentyeight\ respectively. This is the
advantage of this scheme and the expressions of $\beta$ functions
\twentytwo, \twentyfour\ and \twentyfive\ can continuously be used
throughout $2\leq D\leq4$ (one can check they also hold for $D=2$).

In $D=4$ \twentytwo\ and \twentyfour\ precisely
coincide with the result of MS4, the first and the
third of \thirteen.
On the other hand, the form of $\beta_G$ in $D=4$ in both scheme are
completely different; \twentyfive\ and the second of \thirteen.
The non-trivial zero of \twentyfive\ is mapped to infinity in MS4.
The mapping between the present scheme and MS4 is thus somewhat
singular even in $D=4$. In any case, whatever one prefers MS4 or the
conventional scheme in the cutoff regularization, the very fact that
there is no fixed point in the {\it whole coupling constant space\/}
in $D=4$ does not change. In this sense both of them are consistent.

Let us now consider the relation between MS2 and the present scheme.
According to \twentytwo, \twentyfour\ and \twentyfive\ in $2<D<4$
there exist two fixed points. It is easy to see that the critical
exponents at those fixed points are precisely given by the table \ten\
and thus they are nothing but the fixed points (A) and (B) observed in
MS2. When $D\to4$, due to the constant term in \twentytwo\ and
\twentyfour, the fixed points are lost (or go to infinity), while
they survive in \eight. Note that there is no general guarantee that
both of them give a consistent answer for $D\to4$, because MS2 is not
a renormalization in $D=4$ in the sense that Green functions are
not made finite.

The relation between both of schemes should therefore be singular at
$D=4$.
Actually, by identifying $\mu$ in the both schemes, it is easy to see
that
$$
\eqalign{
   &Z_R=Z_R^{\rm MS2}+{2(D-1)\Gamma(D/2)^2\Gamma(3-D/2)\over
                       (2\pi)^{D/2}\Gamma(D)(4-D)},
\cr
   &G_R=
   {G_R^{\rm MS2}\over1-{1\over(D-2)}
    \left[{2\over\pi}-{4(D-1)\Gamma(D/2)^2\Gamma(3-D/2)\over
          (2\pi)^{D/2}\Gamma(D)}\right]G_R^{\rm MS2}},
\cr
   &g_R=g_R^{\rm MS2}+{3\over\pi^2(4-D)}+O((4-D)^0).
\cr
}
\eqn\twentynine
$$
This is the main result of this article:\foot{Note the second
expression is {\it regular\/} at $D=2$} Although the dimensional
continuation of two dimensional MS scheme gives a non-trivial fixed
point at $D=4$, it is an artifact of the dimensional continuation.
In terms of the conventional renormalization scheme in $D=4$
the fixed point corresponds to the infinity and has no physical
relevance.

\section{Conclusion}
As has been shown above, the dimensional continuation of the result of
MS2 to $D\to4$ cannot be used in this model. Our model \one\ seems
of course almost trivial. However for $N\to\infty$ we can eliminate
$\sigma$ using the equation of motion\foot{This may be used to
compute the anomalous dimension of various composite operators,
$(\bar\psi\psi)$, $\square(\bar\psi\psi)$, $(\bar\psi\psi)^3$ etc., by
combing with the fact that $\sigma$ is not renormalized in the leading
order of $1/N$ expansion.}
$$
   \sigma=-{2G\over N}\bar\psi\psi+{4G^2Z\over N}\square(\bar\psi\psi)
   -{8G^3Z^2\over N}\square^2(\bar\psi\psi)
   +{8G^4g\over3}(\bar\psi\psi)^3
   +\cdots,
\eqn\thirteenone
$$
and obtain
$$
\eqalign{
   {\cal L}&=\bar\psi i\gamma^\mu\partial_\mu\psi
             +{G\over N}(\bar\psi\psi)^2
             -{2GZ\over N}\bar\psi\psi\square(\bar\psi\psi)
             +{4G^3Z\over N}\bar\psi\psi\square^2(\bar\psi\psi)
             +{2G^4g\over3N^3}(\bar\psi\psi)^4
\cr
   &\quad-{16G^5gZ\over3N^3}(\bar\psi\psi)^3\square(\bar\psi\psi)
         +{16G^6gZ^2\over N^3}\{
                   (\bar\psi\psi)^3\square^2(\bar\psi\psi)
                   +3[\bar\psi\psi\square(\bar\psi\psi)]^2\}
\cr
   &\quad+O((\bar\psi\psi)^6,\square^3).
\cr
}
\eqn\thirteen
$$
This lagrangian seems highly non-renormalizable, but what the
lagrangian \one\ tells is that the system is renormalizable even in
$D=4$. An Infinite type of UV divergences which appear in the
calculation of \thirteen\ can be removed by the renormalization of
only three parameters, $Z$, $G$ and $g$. In this sense the seemingly
non-renormalizable model \thirteen\ has a predictive power. (This is
of course a trivial statement in view of \one). Note that $Z$, $G$ and
$g$ appear in the coefficients with positive powers in this expansion.
Thus what we considered in this article can be stated: ``Is it possible
to impose the asymptotic safety on \thirteen\ in $D=4$?'' According
to the two dimensional MS scheme, the asymptotic safety puts $Z=g=0$
and the model reduces to four dimensional Gross--Neveu model. We
showed this is not the case.

On the other hand, at $D=3$, i.e., another physical dimension in
$2\leq D\leq4$, there is no contradiction in the above analysis. The
asymptotic safety can therefore be imposed in $D=3$ and the system
reduces to the Gross--Neveu model. This should be so to be consistent
with the fact that the Gross--Neveu model in $D=3$ is renormalizable
[\WIL,\GROSS,\ROS]. This better UV behavior in $D=3$ is clearly
related to the fact that there does not appear new type of UV
divergence in $D=3$ than $D=2$.

Although we considered only a simple (even without gauge symmetry)
model in this article, the fact we observed itself seems independent
on the detail of the model: Though some of fixed points in the whole
coupling constant space appear in a simple form in MS scheme, it
seems in general quite dangerous to continue the result until
the spacetime dimension in which new type of UV divergence (or new
pole in the dimensional regularization) appears.

We thank T. Fujiwara for enlightening discussions. The work of
H. S. is supported in part by Monbusho Grant-in-Aid
Scientific Research No.~07740199 and No.~07304029.

\refout
\bye